\documentclass[useAMS,usenatbib]{mn2e}
\usepackage{graphicx}
\usepackage{fixltx2e}
\usepackage{subfigure}
\usepackage{times}
\usepackage{amsmath}
\usepackage{amsfonts}
\usepackage{url}
\usepackage{placeins}
\usepackage{natbib}
\usepackage{amssymb,amsmath}
\usepackage[normalem]{ulem}
\usepackage{soul}
\usepackage{lipsum}
\usepackage{longtable}
\usepackage{pdflscape}
\usepackage{amsmath}
\usepackage{multirow}
\usepackage{booktabs}

\usepackage{lipsum}

\begin{document}

\title[Quasar HOD at $z \sim 1.5$]{A Halo Occupation Interpretation Of Quasars At $z\sim1.5$ Using Very Small Scale Clustering Information}
\author[Eftekharzadeh et al.]{S. Eftekharzadeh$^{1,2,3}$,  A. D. Myers$^{3}$, E. Kourkchi$^{4}$\\
\\
$^{1}$ Department of Physics and Astronomy, University of Utah, 115 S 1400 E, Salt Lake City, UT 84112\\
$^{2}$ Department of Physics, Southern Methodist University, 3215 Daniel Ave., Dallas, TX 75225\\
$^{3}$ Department of Physics and Astronomy, University of Wyoming, 1000 University Ave., Laramie, WY 82071\\
$^{4}$ Institute for Astronomy, University of Hawaii, 2680 Woodlawn Drive, Honolulu, HI 96822, USA}
\maketitle

\begin{abstract}
We combine the most precise small scale ($< 100\, \rm h^{-1}kpc$) quasar clustering constraints
to date with recent measurements at large scales ($> 1\, \rm h^{-1}Mpc$) 
from the extended Baryon Oscillation Spectroscopic Survey (eBOSS)  
to better constrain the satellite fraction of quasars at $z\sim 1.5$ in the halo occupation formalism. 
We build our Halo Occupation Distribution (HOD) framework based on commonly 
used analytic forms for the one and two-halo terms with two free 
parameters: the minimum halo mass that hosts a central quasar and the fraction of satellite quasars that are within one halo. Inspired by recent studies that propose a steeper density profile for the dark 
matter haloes that host quasars, we 
explore HOD models at kiloparsec scales and best-fit parameters for models with $10\times$ higher concentration parameter. We find that an HOD model with a satellite fraction of $f_{\rm 
sat} = 0.071_{-0.004}^{+0.009}$ and minimum mass of $\rm M_{m} = 2.31_{-0.38}^{+0.41} \times 10^{12}\, \, \rm h^{-1} M_{\odot}$ 
for the host dark matter 
haloes best describes quasar clustering (on all scales) at $z \sim 1.5$. Our results are marginally inconsistent with earlier work that studied brighter quasars, hinting at a luminosity-dependence to the one-halo term.

\end{abstract}


\section{Introduction}
\label{intro}
The advent of the first large, homogeneous surveys of the extragalactic sky prompted
cosmologists to begin to think about galaxies as discrete points embedded in 
statistical structures \citep[e.g.][]{nsz62,ns74}. These structures could
be characterized in terms of their size, the distribution of the discrete points
that occupied them, and their distribution across the wider Universe
 \citep[see, e.g. ][for a review]{cs02}. As it became increasingly clear
 that dark matter dominated the mass budget in galaxies \citep[e.g.][]{Roo72,Ost74,Rub78}, it became more natural
 to think of occupation statistics in terms of the virialized haloes that host 
 luminous tracers \citep[e.g.][]{Whi78}.
This line of reasoning ultimately led to empirical approaches that describe how 
cosmological tracers occupy underlying dark matter structures, such as the Halo 
Occupation Distribution framework \citep[HOD; e.g., ][and references therein]{bw02,zh07}.
Parameterizations of the HOD typically consist of of a two-halo term, which characterizes how haloes of a certain
mass cluster around each other, and a one-halo term that relates galaxy and dark matter statistics through the
probability that a halo of a given mass contains a number of galaxies of a given
type. 

The key observables that are used for constraining HOD descriptions in the common formalism
are the number density and the clustering of a given tracer population. 
As such, the HOD framework has now been successfully used to model galaxy
clustering measurements for a wide range of redshifts and galaxy
types \citep[see, e.g.,][]{pmn04,ab05,pwnp09,zhe09,si09,st11,zeh11,cou12,guo14,bh18}. 

Quasars, the most luminous of the Active Galactic Nuclei, are driven by accreting
supermassive black holes at the centers of galaxies. It is now well-established that the centers of most galaxies 
contain a supermassive black hole \citep[e.g.][]{Kor95}, and that the evolution of active quasars and inactive 
galaxies is interrelated \citep[e.g.][]{Kau00}. It is therefore reasonable to think of quasars simply as a biased tracer
of certain types of galaxies that should also, in theory, be empirically describable using HOD statistics.

In the wake of large spectroscopic surveys such as the 2dF \citep[][]{2dFsurv} and SDSS \citep[e.g.,][]{fu96,gu98,yo00,va01,st02,str02,te04,we04,yi04a,yi04b,bo04,ro05,wi05}, the clustering of quasars has been measured at a range of redshifts
\citep[e.g.,][]{cr04,po04,my06,my07,my07b,she07,she09a,ro09,ric12,wh12,ric13,ro13,ef15,la17}.
Typically, these studies focus on the large-scale clustering of quasars via the two-point correlation function, which 
constrains the ``two-halo term'' that describes how ``central'' dark matter haloes cluster around each other. The consensus is that, 
at most redshifts, quasars occupy central haloes of masses a few times $10^{12}\,h^{-1} {\rm M_{\odot}}$. 

Probing how quasars are distributed within haloes---the so-called ``one-halo'' term---is trickier, however. As 
quasars occupy massive haloes, they are rare in general. Further, the haloes that host quasars, particularly at 
high redshift \citep[e.g.,][]{wh12,ef15}, are on the steeply falling part of the halo mass function.
This implies that instances of two quasars occupying a single halo at high redshift may be very rare indeed. 
Finding close pairs of quasars is complicated further by the fact that most large spectroscopic surveys use 
fiber-fed multi-object spectrographs. Such surveys can have restrictions on how closely fibers can be placed 
together on the sky, which has prompted follow-up surveys of close quasar pairs using long-slit spectrographs 
\citep[e.g.,][]{Hen06,my08,Hen10,ko12,ef17}. These long-slit surveys, in combination with the large-scale two-point
correlation function, have been used to constrain the one-halo term for quasars via clustering measurements over 
a wide redshift range ($z \sim$ 0.5--3) and scale \citep[][]{ric12,ko12,sh13}.

Recent measurements of quasar clustering have used a number of different assumptions for the overall form
of the quasar HOD. For instance \citet[][henceforth KO12]{ko12} model both the distribution of satellite and
central quasars using Gaussians, ultimately expressing the HOD using two-to-three fitting  
parameters. On the other hand, \citet{zh07}, \citet{ric12,ric13} and \citet{sh13} use a model that combines a power-law with
a Gaussian, which requires five-to-six fitting parameters. These choices, however, seem to have little
power to constrain the one-halo term of the quasar HOD as KO12 and \citet{sh13} derive consistent 
satellite fractions using their two different HOD parameterizations. Much of this degeneracy regarding the form of the quasar 
HOD is driven by sizeable uncertainties in the parameters that are fit to model quasar clustering on small scales. It is likely, 
then, that much larger samples of quasar pairs with small separations, or alternative approaches to deriving the 
Mean Occupation Function of quasars \citep[e.g.][]{Cha13}, will be needed to probe the overall statistics of how
quasars occupy individual haloes.

Despite the range of possible forms for the Mean Occupation Function of quasars, it remains important to provide empirical 
constraints on the quasar HOD. Large surveys such as the quasar component \citep{my15} of the
{\em extended Baryon Oscillation Spectroscopic Survey} \citep{daw16} and surveys with the 
{\em Dark Energy Spectroscopic Instrument} \citep{DESI1,DESI2} are beginning to use 
quasars to constrain the cosmological world model at moderate redshift via redshift-space distortions and the Baryon 
Acoustic Oscillation scale. Sophisticated simulations are required to model these cosmological constraints, which require 
an assumed form for the quasar HOD on small scales \citep[e.g.][]{rod17}. Recently, we assembled by far the largest sample of 
quasar pairs that can be used to probe quasar clustering on scales of a few dozen kiloparsecs, well into the one-halo 
regime \citep[][]{ef17}. In this paper, we combine the sample of \citet{ef17} with other clustering results on larger 
scales \citep[e.g.][]{ko12,la17} to provide the best current constraints on the quasar HOD.

This paper is structured as follows: $\S$\ref{dat} summarizes the properties of 
the samples that are used in our small- and large-scale clustering measurements. The measurements 
themselves are detailed in $\S$\ref{cls}, the modeling approach and the chosen 
parameterization is described in $\S$\ref{hodmod}, we discuss our main results in $\S$\ref{res},
and present our conclusions in $\S$\ref{con}. We adopt a $\Lambda$CDM cosmological model 
with matter and dark energy and baryon density densities of $\Omega_{m}=0.307$, 
$\Omega_{\Lambda}=0.693$, and $\Omega_{b}=0.045$, a Hubble parameter of $h=0.678$, 
amplitude of matter fluctuations $\sigma_{8}=0.814$, and the slope of the 
initial power spectrum $n_s=0.968$, consistent with 
\citet{planck15}. All distances quoted throughout the paper are in comoving 
coordinates unless noted otherwise. We denote proper coordinates by adding ``p'' to the 
distance units (i.e., $h^{-1}{\rm pkpc}$ or $h^{-1} {\rm pMpc}$). 

\begin{figure*}
    \centering
    \begin{subfigure}{
        \centering
        \includegraphics[angle=0,scale=0.29]{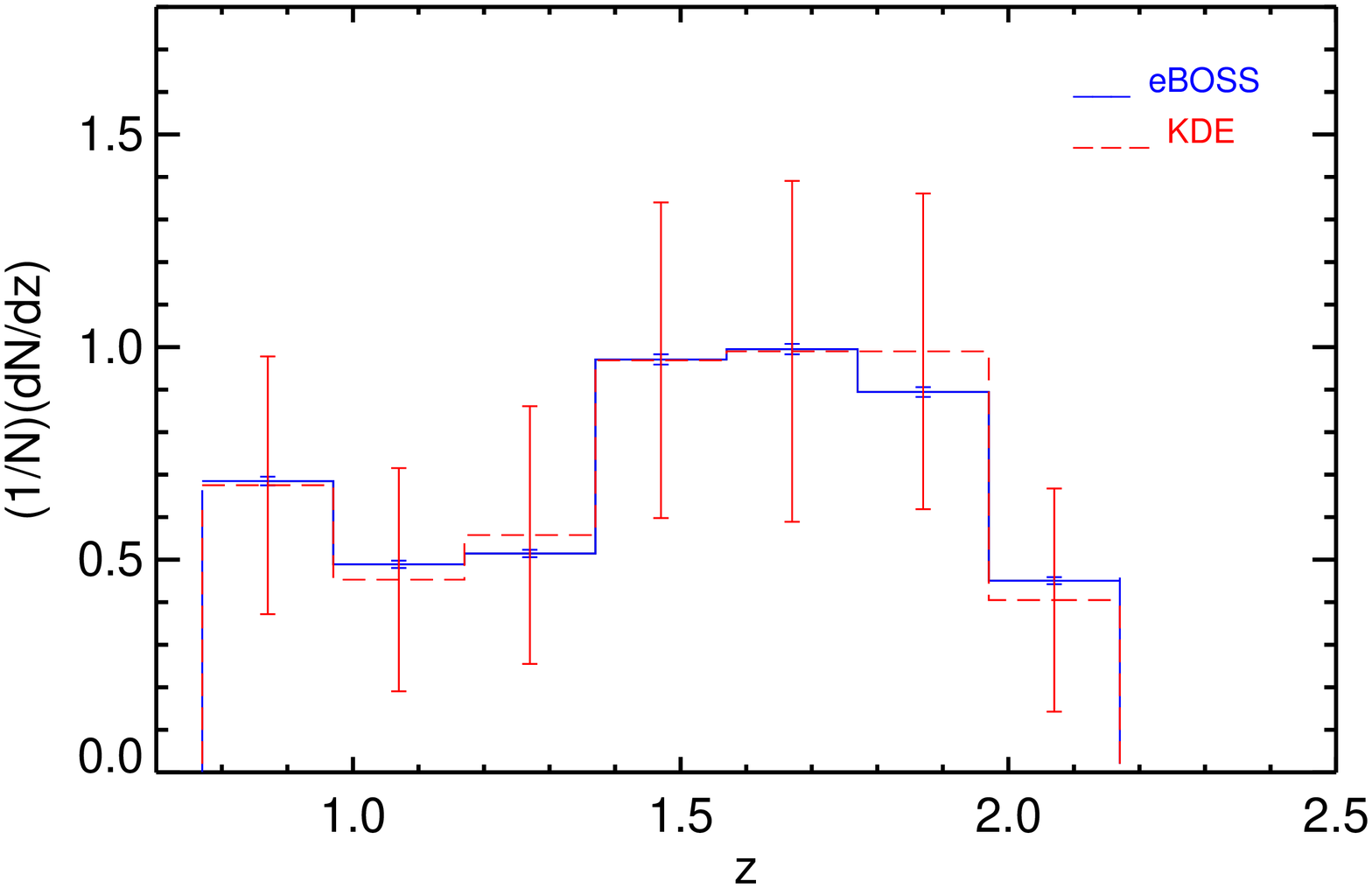}
        \includegraphics[angle=0,scale=0.29]{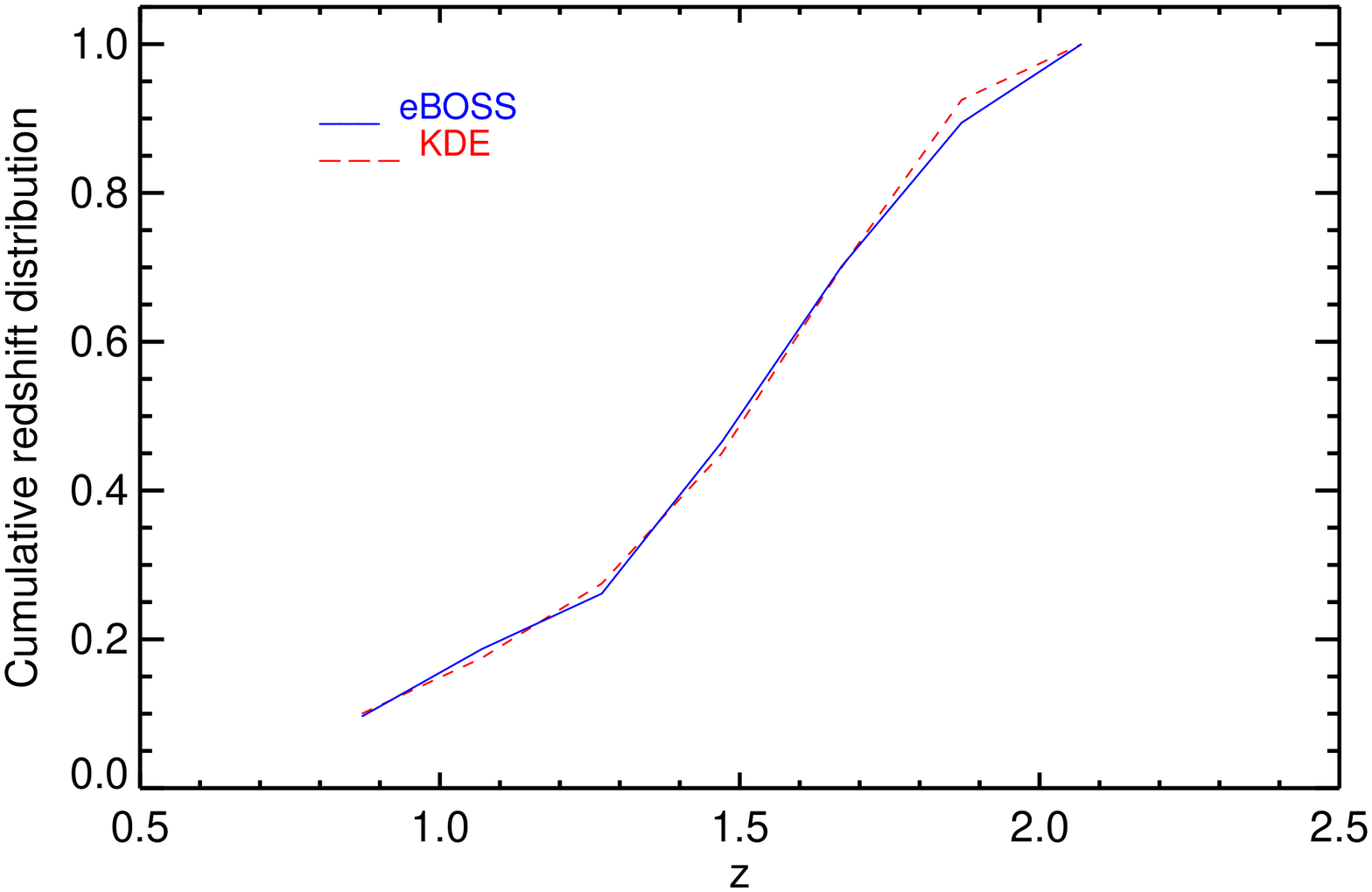}}
        \caption{Left: Normalized redshift distribution of the KDE-complete sample of ($<7 \arcsec$) pairs (dashed red line) and the down-sampled redshift distribution 
        of the eBOSS quasars (solid blue line). The ``downsampling'' process was performed to match the redshift distribution of the KDE-complete and eBOSS quasars in order to
         to jointly model their clustering measurements (see $\S$\ref{dat}). Right: the cumulative distribution function (CDF) of our KDE and eBOSS samples.} \label{fig1:nz-cdf}
        \end{subfigure}

\end{figure*}

\begin{table*}
\centering
\begin{tabular}{cccccccc}

\hline
\hline
$z_{\rm min}$   &$z_{\rm mid}$ &$z_{\rm max}$& $(1/N_{\rm eBOSS})~dN/dz$ & Error & $(1/N_{\rm KDE})~dN/dz$& Error\\

\hline

0.77 & 0.87 & 0.97 & 0.685 & 0.010 & 0.676 & 0.303 \\
0.97 & 1.07 & 1.17 & 0.489 & 0.009 & 0.455 & 0.262 \\
1.17 & 1.27 & 1.37 & 0.515 & 0.009 & 0.558 & 0.303 \\
1.37 & 1.47 & 1.57 & 0.971 & 0.012 & 0.969 & 0.371 \\
1.58 & 1.67 & 1.77 & 0.994 & 0.012 & 0.990 & 0.401 \\
1.77 & 1.87 & 1.97 & 0.895 & 0.012 & 0.990 & 0.371 \\
1.97 & 2.07 & 2.17 & 0.451 & 0.008 & 0.405 & 0.262 \\

\hline
\end{tabular}
\caption{ Normalized distribution of the spectroscopic redshifts of quasars in eBOSS (4th column) and the KDE-complete sample of close pairs (6th column) in the 
redshift bins depicted in Fig.\,\ref{fig1:nz-cdf}.}\label{tab:dndz}

\end{table*}

\begin{figure*}
    \centering
    \begin{subfigure}{
        \centering
        \includegraphics[angle=0,scale=0.27]{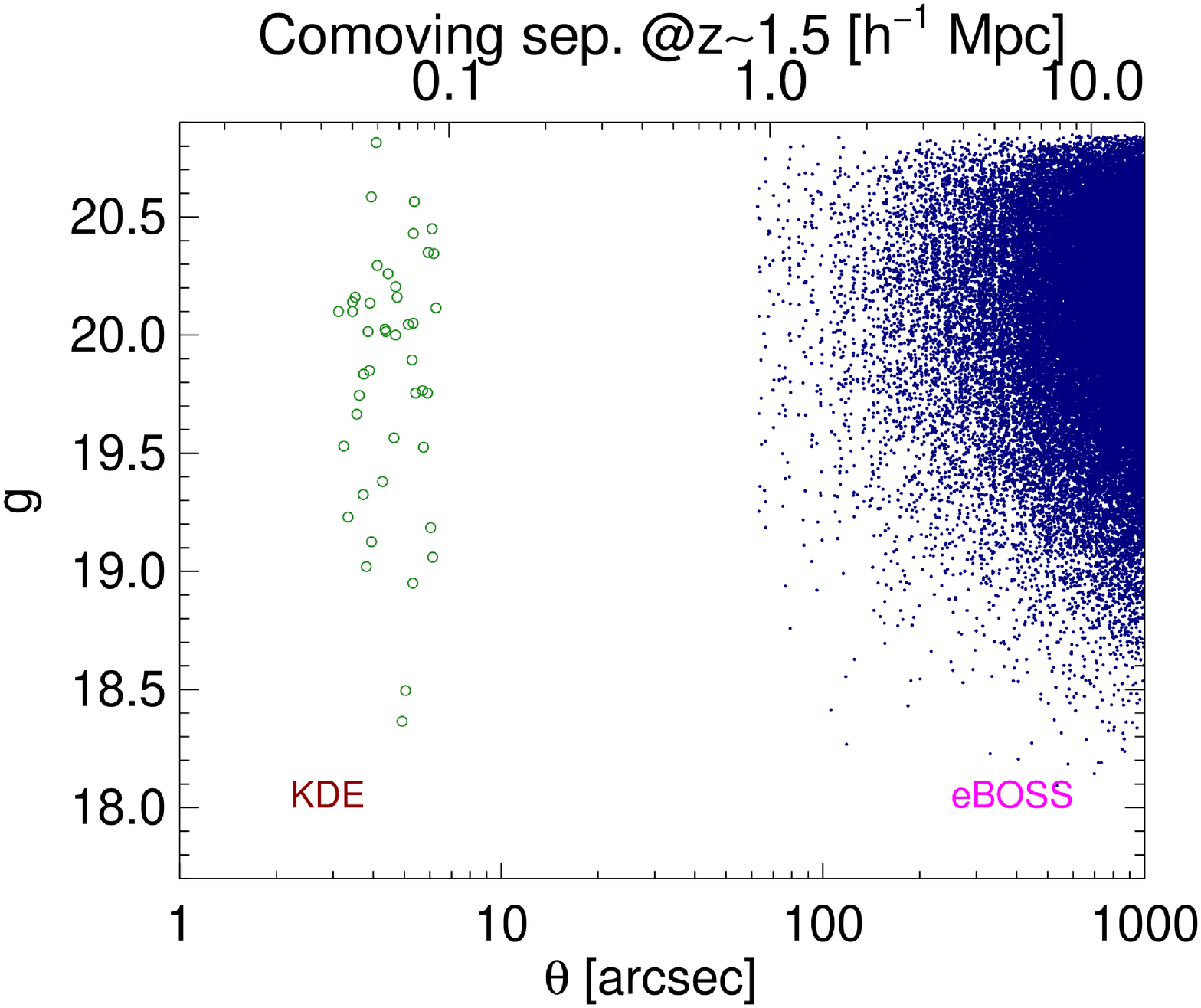}
        \includegraphics[angle=0,scale=0.27]{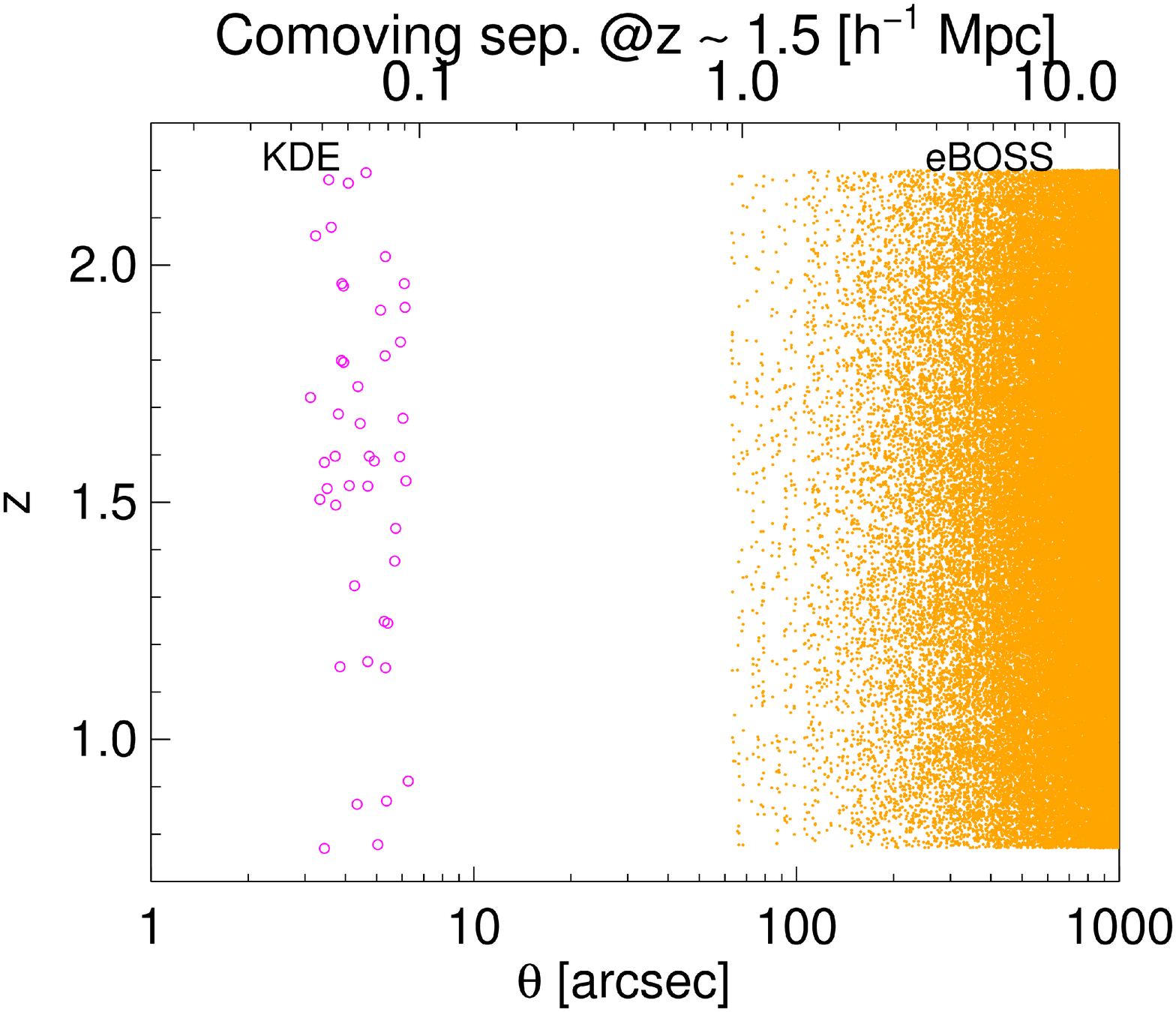}}
        \caption{Angular separation of quasars in each of the pairs that are 
used for the small and large-scale clustering measurement as a function of their 
$g$-magnitudes (left) and their redshifts (right). Blue dots and green circles in 
the left-hand panel and orange dots and pink circles in the right-hand panel each represent one 
quasar {\em pair}. The open circles in both panels represent the 44 ($< 7 
\arcsec$) pairs drawn from our KDE-complete ``parent sample'' of  243{,}110 quasar 
candidates \citep[see][for details]{ef17}. The blue and orange dots in the left-hand and 
right-hand panels are the quasar pairs in a sample of 40{,}821 $g\le 20.85$ quasars 
from eBOSS. For consistency, we applied the same magnitude and redshift limits to both samples. 
The comoving transverse separation at the average redshift of both samples ($z\sim 
1.5$) is included as the top axis in each panel.}\label{fig2:gz}
    \end{subfigure}
\end{figure*}

\section{Data}\label{dat}
 We use two independently compiled samples of confirmed quasars for clustering 
measurements at kpc and Mpc scales. The small-scale clustering measurement used in this work is drawn from a complete sample of 
$g<20.85$ confirmed quasars at $z \sim 1.5$ \citep[see][]{ef17} and the large-scale clustering measurement is 
determined using a sample of $g<20.85$ quasars from the extended Baryon 
Oscillation Spectroscopic Survey \citep[eBOSS; DR14][]{daw16,dr14}, also at $z \sim 1.5$. We apply a number of adjustments prior to measuring the two-point correlation function and combine these two measurements into a single constraint that spans comoving distances of $\sim 0.01$--$100 \, h^{-1}{\rm Mpc}$. 
In this section of the paper, we summarize the compilation process and properties of the two samples.

\subsection{KDE-complete sample of close pairs}\label{sdat}

On small scales, we study the ``KDE-complete'' sample of 47 spectroscopically confirmed ``binary'' quasars from \citet{ef17}, which have 
angular separations of $2.9\arcsec \leq \theta \leq 7.7\arcsec$ over a redshift range of $0.43 \leq z \leq 2.2$.
The target sample from which these 47 spectroscopically confirmed ``binaries'' was drawn is 1{,}172{,}157 high-probability quasar candidates compiled by applying
a Kernel Density Estimation technique \citep[KDE;][]{ric04} to all the point sources in Data Release 6 of the SDSS
\citep[][]{admc08} imaging data down to $i=21.3$. A total of 290{,}694 quasar candidates at $0.43 \leq z \leq 2.2$ was selected via a non-parametric Bayesian classifier with $\sim 92.7\%$ efficiency \citep[see][and references therein for details]{ef17}. A carefully designed long-slit spectroscopic campaign observed a homogeneous subsample of these pair candidates over the course of three years \citep[see Table 1 of ][]{ef17}. 

Of the 47 pairs in the KDE-complete sample \citep[presented in Figure 4 and Table 5 of][]{ef17}, 44 reside in the selected redshift range for our current study ($0.8 \lesssim z \lesssim 2.2$ with $\bar z\sim 1.5$). The comoving separations of quasars in these pairs span $43.3 \lesssim r_p \lesssim 92.3 \, h^{-1}{\rm kpc}$. We use this sample to make our sub-Mpc quasar clustering measurement, creating a random catalog as outlined in section 3.1 of \citet{ef17}.

\subsection{eBOSS quasars with $g<20.85$}\label{ldat}

To produce a ``Mpc-scale'' clustering measurement, we use a large sample of spectroscopically confirmed quasars from the extended Baryon Oscillation Spectroscopic Survey \citep[eBOSS;][]{daw16} targeted as described in \citet{my15}.  
We begin with a set of 116{,}866 eBOSS quasars \citep[released in SDSS-DR14;][]{dr14} and its associated $\sim 44 \times$ larger random catalog. The random catalog models variations in the completeness of eBOSS quasars across the survey footprint in order to reproduce the number density and redshift distribution of the targets \citep[see, e.g.][]{sw08b,la17}.     

The eBOSS random catalogs that we adopt are similar to those used in \citet{la17} and \citet{rod17}, which include weights to correct for 
incompleteness in identifying quasars \citep[see, e.g., ][]{zh18,za18}.
Corrections include a depth-dependent systematic weight ($w_{sys}:$ {\tt WEIGHT\_SYSTOT}), a weight to account for redshift failures ($w_{zf}:$ {\tt WEIGHT\_NOZ}), and a fiber collision ($w_{cp}:$ {\tt WEIGHT\_CP}) term \citep[see][for more details regarding how weights are determined]{an12,ross12,bp17}. The total weight for each quasar is ultimately defined as $w_{\rm qso} = w_{\sc fkp}w_{sys}(w_{zf}+w_{cp}-1)$ where $w_{\sc fkp}:$ ({\tt WEIGHT\_FKP}) is designed to optimally estimate the two point correlation function \citep[see][]{fkp94}.

Due to the size of the ferrules that support the fibers that feed the eBOSS spectrographs, eBOSS cannot place fibers closer than $62 \arcsec$, except in regions where multiple spectroscopic plates overlap \citep{daw16}. To measure clustering on small scales, it is critical to be complete as the interplay between the fiber collision radius and the {\it unknown} small-scale clustering of the targets is impossible to perfectly reconstruct because quasar samples are sparse enough that the number of collided pairs is very small \citep[e.g.,][]{rod17}. 
As the $62 \arcsec$ eBOSS angular separation limit is equivalent to a comoving separation of $0.913\, h^{-1}{\rm Mpc}$ at $z\sim 1.5$ , throughout the current paper we set the first bin of comoving separation at which we start to measure our ``Mpc-scale'' two-point correlation function (2PCF), to be $\sim 1.0 \, h^{-1}{\rm Mpc}$ to ensure that we are on scales larger than the fiber-collision limit. See \citet{gu12} and \citet{ha17} for detailed analyses of the efficiency of fiber collision corrections at small scales.

\subsection{Matching quasar samples in luminosity and redshift}\label{matchdat}
A total of 88{,}764 of the quasars discussed in \S\ref{ldat} fall in the redshift range $0.8\lesssim z \lesssim 2.2$, of which 40{,}821 also have $g<20.85$, matching the redshift and brightness limits for the quasar pairs in the KDE-complete sample discussed in \S\ref{sdat}. Maintaining a similar redshift and luminosity range for quasars in both our ``KDE-complete'' and ``Mpc-scale'' samples would be ideal to guarantee that our clustering measurement studies a consistent population of quasars across all scales, although any luminosity evolution in quasar clustering should be small in the regime we consider \citep[e.g.][]{she07,All11,sh13,All14,ef15,mg16}. 

We match the distributions by {\it downsampling} the much larger eBOSS quasar sample in over-populated redshift bins to that of the KDE-complete sample of close pairs, by first determining the over population at each bin of redshift and then randomly removing the {\it extra} number of targets from that redshift bin. Figure \ref{fig1:nz-cdf} shows the normalized redshift distribution of the KDE-complete sample of ($<7 \arcsec$) pairs and the down-sampled and matched redshift distribution of the eBOSS quasars. The right-hand-side of this figure illustrates the cumulative probability distribution function (CDF) of the KDE and eBOSS samples. The two-sample Kolmogorov-Smirnov test shows that the probability that these two samples are drawn from the same distribution is $>$\,$86\%$. The errors on the redshift distribution of the KDE sample are Poisson errors to show that the match is well within the one-sigma uncertainty. We provide the resulting matched redshift distributions of the KDE-complete sample of 44 pairs and the 33{,}245 eBOSS $g<20.85$ quasars in Table\,\ref{tab:dndz}. 
  
Figure \ref{fig2:gz} shows the angular separation of quasars in each of the pairs that we 
use for the small- and large-scale clustering measurement as a function of their 
$g$-magnitudes and their redshifts. 
As is evident in Fig. \ref{fig2:gz}, both sets of pairs from the small and large angular separations 
exhibit similar distributions in magnitude and redshift. 

As noted in \S\ref{ldat}, fiber-collision-effects
impose a lower limit of $62\arcsec$ for the 
angular separation of the quasar pairs. This produces a $\sim 55\arcsec$ gap between 
the close pairs from the KDE-complete sample at $7\arcsec$ and the eBOSS 
pairs.

\section{Clustering Measurements}\label{cls}
In this section, we summarize 
how we use the samples outlined in \S\ref{dat} to make quasar clustering 
measurements at $z\sim 1.5$ on kpc and Mpc scales. 

\subsection{kpc-scale clustering}\label{kpcls}
The method we adopt for measuring the volume-averaged correlation function in real space with sparse samples of close pairs is 
discussed in \citet{ef17} and \citet{Hen06}. Fig. 2 of \citet{ef17} shows how the sample of 47 pairs with angular separations of 
$<7\arcsec$ that we outlined in \S\ref{sdat} constitute a complete sample and how they populate the redshift-physical-separation plane. 
We follow the method depicted in Fig. 2 of \citet{ef17} but apply an additional redshift cut of 
$0.8 \lesssim z \lesssim 2.2$ to mirror the eBOSS sample that we will use to measure quasar clustering on Mpc-scales. 
This results in a sample of 44 pairs that are complete across a range of proper scales,
which we convert to comoving coordinates.

The ``KDE-complete`` sample discussed in \citet{ef17} was designed to be complete in terms of spectroscopic follow-up on {\em proper} scales but we conduct our clustering measurement on {\em comoving} scales. To transform our measurements to comoving scales, for which the sample may not be complete, we reconstructed the separation-redshift ($R-z$) plane for our sample of quasar pairs (e.g., the filled orange circles in Fig. 2 of \citealt{ef17}) and checked that our redshift and comoving scales of interest do not include any pair with an unconfirmed spectroscopic redshift. Transforming the $R-z$ plane from proper to comoving coordinates we found that one additional pair would require spectroscopic follow-up to ensure that our sample of interest is complete on comoving scales. Including this pair would increase the number of pairs in the bin of comoving separation of $77.1<r_p<92.3 \, h^{-1} {\rm kpc}$ that we discuss in \S\ref{res} from 13 to 14 pairs. Effectively, by ignoring this pair, we are adding an additional 6 to 8 per cent to the uncertainties that we shall discuss in \S\ref{res} and that we summarize in Table \ref{tab:res}. Given our fitting errors, this change is too small to have any noticeable effect in the shape of the best-fit models and consequently the values of the best-fit parameters. We therefore proceed as if our sample is spectroscopically complete.

We show the measured volume-averaged correlation function ($\rm \bar W_p$) for the resulting 44 pairs of quasars in 4 bins of separation
in Fig.\ \ref{fig4:bfit}. As each quasar pair is only counted once (i.e.\ is independent), we adopt Poisson errors from \citet{geh86} for the 
uncertainty of the measured $\rm \bar W_p$. The volume-averaged correlation function was measured for 6, 13,12 and 13 quasar pairs in four bins of separation at comoving average separations of 48, 58, 70 and 85 $h^{-1} {\rm kpc}$, respectively. Similarly to the measurement presented in \citet{ef17}, we also measure $\rm \bar W_p$ for the full bin of 43.4 to 92.3 $\rm h^{-1} kpc$ (shown as a single bin at $\bar r_p = 67.8 \, h^{-1} {\rm kpc}$ in the right-hand panel of Fig.\,\ref{fig4:bfit}). The process through which the {\it expected} Quasar-Random pairs (i.e. the random catalog needed to calibrate clustering), $\langle QR \rangle$, is calculated is detailed in \citet{ef17}.

The volume-averaged correlation function, $\rm \bar W_p$, is useful in characterizing quasar clustering on small scales, as close quasar pairs
are scarce. Quasar clustering is more typically {\em modeled}, however using the projected correlation function 
(i.e., $w_p(r_p)=2\int_{0}^{\infty} d\pi \xi(\pi, r_p)$). We therefore convert $\rm \bar W_p$ to the 
equivalent $w_p(r_p)$ to make it easier to compare our measurements to HOD models. 
We use the approximation:

\begin{equation}
 \bar W_p(r_p) \sim \frac{1}{N_{\rm qso}} \int_{z_{\rm min}}^{z_{\rm max}} dz \frac{dV_c}{dz} n(z) \frac{1}{v_z} \int_{0}^{v_z} d\pi \xi(\sqrt{\left(r_p^2+\pi^2\right)},z),
\label{eqn:one}
\end{equation}

where $n(z)$ is the comoving number density of quasars in bins of redshift, $v_z \equiv v_{\rm max}(1+z)/H(z)$, $v_{\rm max}= 2000 \, \rm km\,s^{-1}$, $H(z)$ is the expansion rate at redshift $z$, $N_{\rm qso} \sim \int_{z_{\rm min}}^{z_{\rm max}} dz \frac{dV_c}{dz} n(z)$ and $\int_{0}^{v_z} d\pi \xi(\sqrt{\left(r_p^2+\pi^2 \right)},z)$ is essentially $w_p(r_p, v_z \rightarrow \infty)$. We then solved the above $W_p(r_p)=f\times w_p(r_p)$ equation for $w_p$ using the calculated $\rm W_p$ in each bin of comoving separation and across the {\it full} redshift range. The factor $f$, here, represents the portion of the right-hand side of the equation that approximates the volume of a shell in redshift space. This shell encompasses the quasar pairs that have the specified comoving separation in the $r_p$ bin. See \S3.2 of \citet{ef17} for additional details.

\subsection{Mpc-scale clustering}\label{mpcls}

\begin{figure*}
    \centering
        \includegraphics[angle=0,scale=0.6]{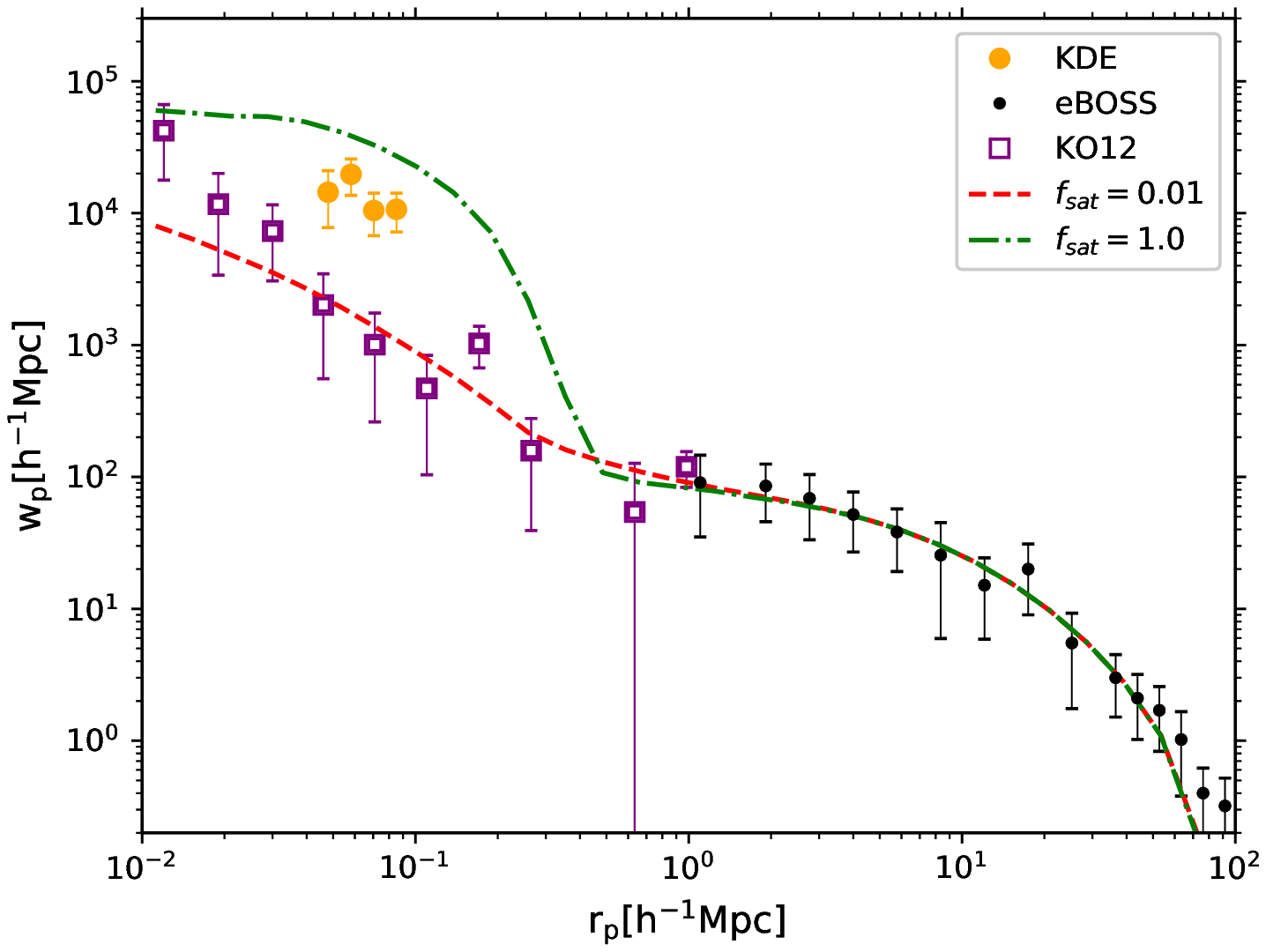}
         \caption{A depiction of the measurements that are used in this study. 
The red dashed and the green dotted-dashed curves are the predicted HOD model 
for satellite fractions of 0.01 and 1.0, respectively, and are only included here to guide 
the eye. The purple squares are the measurement of \citet[][referred 
to as KO12 in the plots and throughout the text]{ko12} and the orange circles are the 
correlation function of $<7\arcsec$ pairs drawn from the KDE-selected parent 
sample detailed in \S\ref{dat}.}\label{fig3:obs}
\end{figure*}

\begin{figure*}
    \centering
    \begin{subfigure}{
     \centering
       \includegraphics[angle=0,scale=0.6]{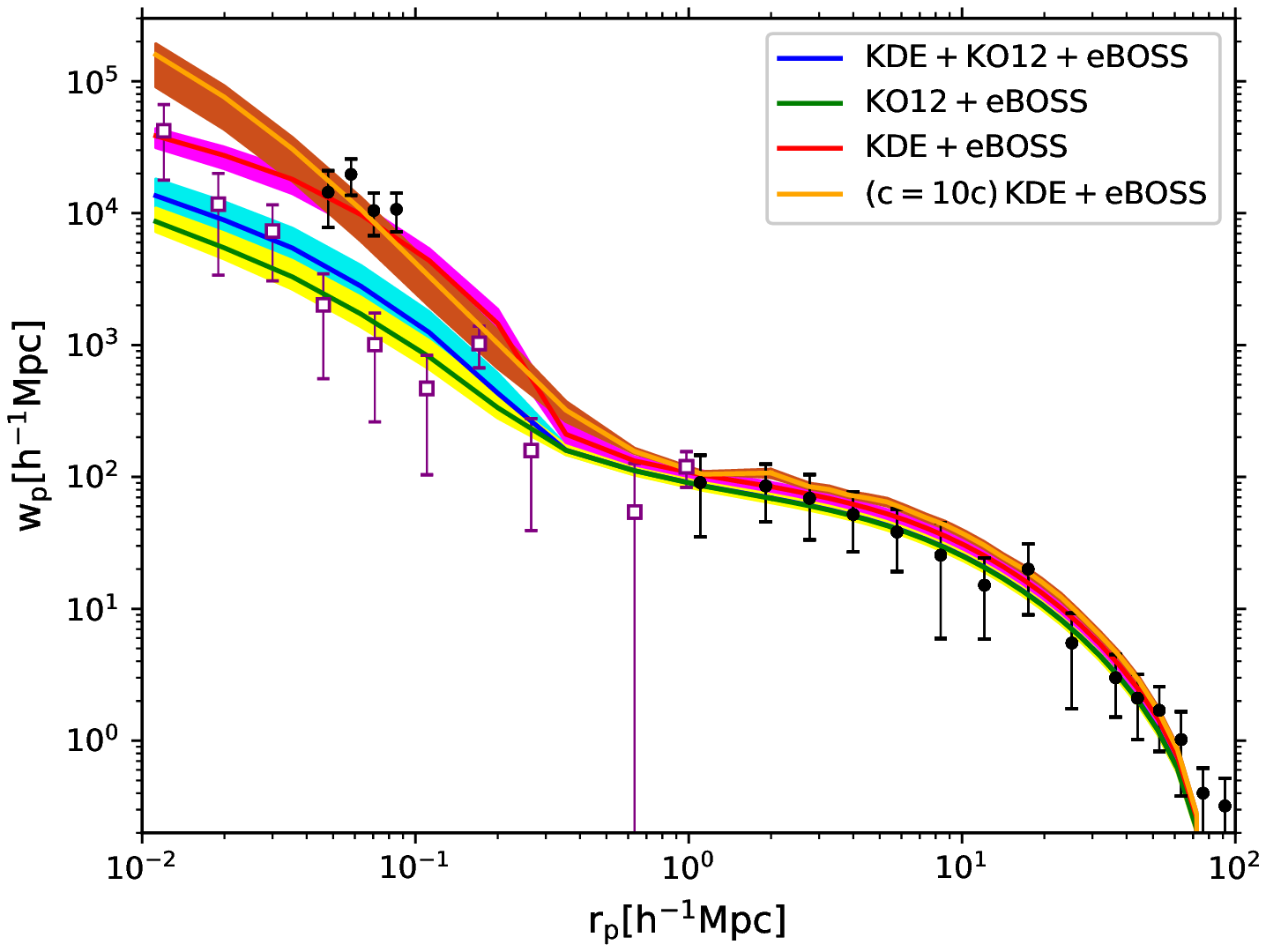}
       \includegraphics[angle=0,scale=0.6]{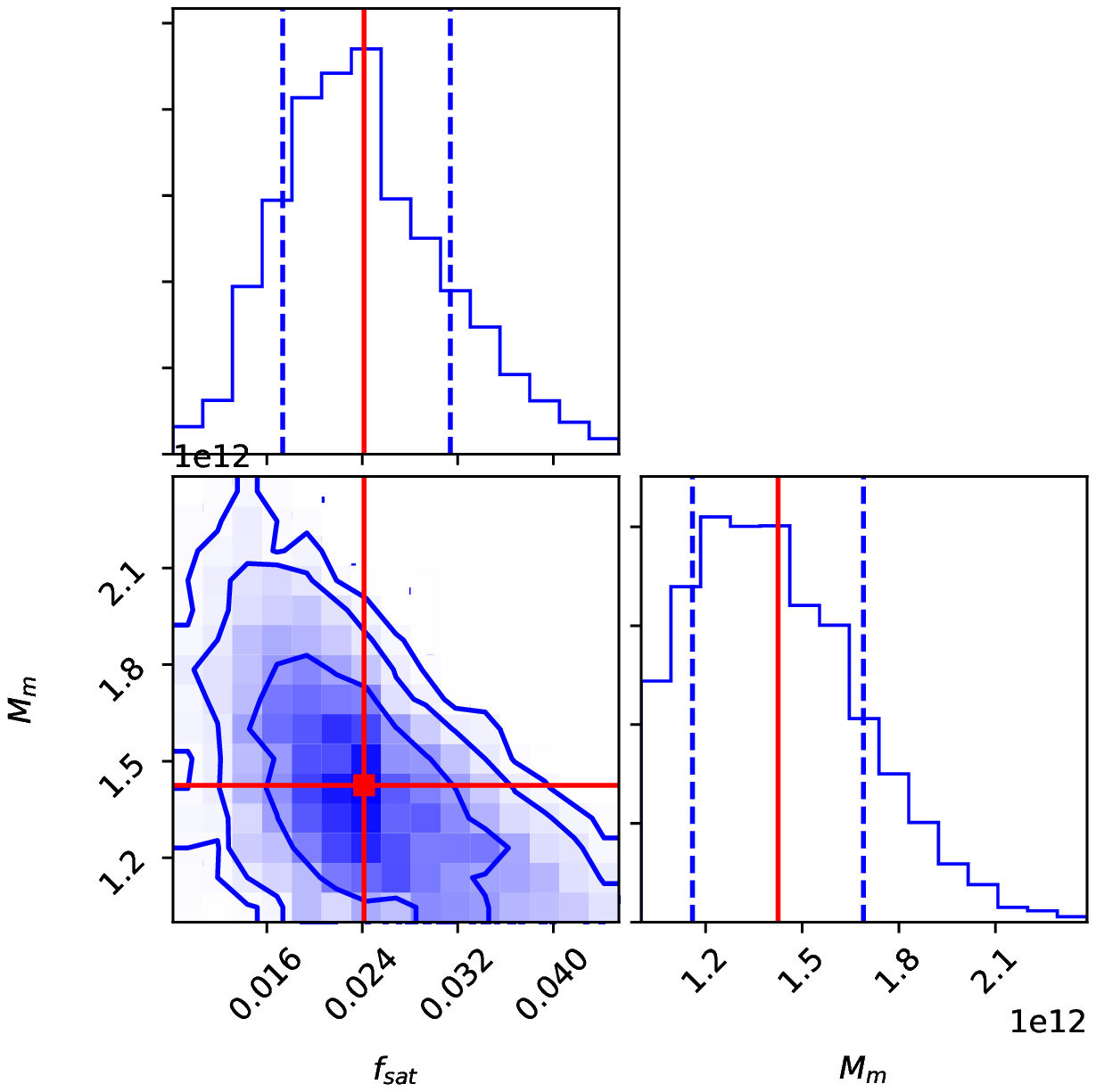}}
    \caption{Left: Our HOD prediction of the projected correlation function that 
best fits the observed correlation function from a combination of KDE, KO12 and eBOSS quasar samples. The fitting results are summarized in 
Table \ref{tab:res}. Shaded envelopes around each best-fit model represent the uncertainty for the shape of the HOD model, corresponding
to 68\% of the density of samples being enclosed by the inner contour in the right-hand panel. Right: The result of 
the MCMC fit to the projected two point correlation function for one of our models. The fit is performed over scales of
$\sim\,0.01$--$100\,h^{-1} {\rm Mpc}$ and produces the solid blue curve in the left-hand panel. 
Each of the four best-fit models in the left-hand panel 
has a corresponding corner plot similar to the right-hand panel, but we only show one such corner plot for brevity.
The diagonal panels show the probability distribution functions (PDFs) for the two 
fitting parameters (i.e., $f_{\rm sat}$ and $M_{\rm min}$). 
The off-diagonal panels show the density contours of the chosen 
sets of HOD parameters corresponding to enclosed 68\%, 95\%, and 99\% of the density of samples. }\label{fig4:bfit}
    \end{subfigure}
\end{figure*}

We calculate the Mpc-scale section of the correlation function using the sample of 
eBOSS quasars described in \S\ref{matchdat} that have been down-sampled to match the number density of the KDE parent sample 
with which the kpc-scale correlation function is measured ($\sim6.5 \times 10^{-6} \, h^3 {\rm Mpc}^{-3}$). The corresponding 
random catalog for the down-sampled quasar catalog is created by keeping 
track of the fraction ($f$) of objects that are brighter than $g=20.85$ in each bin of 
redshift. For each object in the original random catalog and its 
assigned redshift, if a random number between 0 and 1 is less than $f$, then that object is retained,
otherwise it is discarded. We calculate $w_p(r_p)$ using the estimator of \citet{ls93}. 
As discussed in $\S$\ref{ldat}, the quasars that are counted in the Quasar-Random and Quasar-Quasar pairs, have an associated weight $w_{\rm qso}$, which we apply to the pair counts. 
The measured correlation function in bins of comoving separation across $1<r_p<100 \, h^{-1} {\rm Mpc}$ are shown with filled black circles in 
Fig.\,\ref{fig3:obs} and Fig.\,\ref{fig4:bfit}. The error bars are calculated through a jackknife resampling \citep[see, e.g., eqn. 4 of][]{ef15}.

\section{HOD Modelling}\label{hodmod}

One important application of measuring the real-space correlation function of a 
given population is to provide the statistics of how those objects populate 
individual dark matter haloes. The Halo Occupation Distribution framework provides 
the average number of objects residing within one halo (i.e., $\langle 
N(M)\rangle$) as well as an analytical form for 
their distribution in each halo under the assumption that $\langle 
N(M)\rangle$ is a function of the halo mass \citep{pea00,sel00,sc01,bw02}. Variations of this modelling 
approach have been used for interpreting the measured correlation function of 
Active Galactic Nuclei (AGNs) and quasars in recent years \citep[e.g., 
][]{po04,coil04,Aba05,coi06,coi07,coi09,mi11,ric12,ko12,kru12,ric13,sh13,Coi16,coi17}.  

Although several studies have attempted to constrain HOD parameters for quasars on both 
large and small scales, measured quasar correlation functions on halo scales ($<1 \, h^{-1} {\rm Mpc}$)
are rare, and so HOD modeling of the correlation function on these scales has been limited. In this paper, 
we adopt an HOD approach 
similar to KO12, which modeled quasar clustering over a 
similar scale and redshift regime to our current work.  

The projected two-point correlation function can be modeled using the matter 
power spectrum \citep[see, e.g., ][]{cs02}:
\begin{equation}
w_{p}(r_p) = \int^{\infty}_{0} \frac{k dk}{2\pi} P(k) J_0(k r_p),
\end{equation}
where $J_0$ is the zeroth order of the Bessel function of the first kind. The 
power spectrum $P(k)$ can be separated into two independent {\it one-} and {\it 
two-halo} terms: 
\begin{equation}
  P(k) = P_{1h}(k)+P_{2h}(k).
\end{equation}
The two-halo term can be simply obtained by  
\begin{equation}
P_{2h}(k) = b^2 P_{\rm lin}(k),
\end{equation}
where the linear matter power spectrum, $P_{\rm lin}$ is computed using 
the fitting form from \citet{eh99} with our chosen cosmological parameters and $b$ is 
the bias parameter that can be modeled as:

\begin{equation}
b = \frac{\int b_h(M) \frac{dn}{dM} dM}{\int \frac{dn}{dM} dM}.
\end{equation}

Where $b$ is the mass-averaged bias of the haloes above a characteristic minimum mass, weighted by the halo abundance $dn/dM$ \citep[see, e.g., Eqn.\,9 of][]{ef15}. We take $dn/dM$ from the halo mass function of \citet{tin08} at the mean redshift of our quasar sample.
Following a commonly assumed mass-dependence, we model the mean number of 
occupying quasars in individual haloes as a three-parameter Gaussian: 

\begin{equation}
\langle N(M) \rangle = f_{N} \times \frac{1}{\sqrt{\left(2\pi \right)}\Delta_m} \exp\left[-\frac{\ln^2(M/M_m)}{2\Delta^{2}_m}\right],
\end{equation}

where $M_m$ is the minimum mass for a dark matter halo that can host a quasar 
and $\Delta_m$ is the width of the initial mass distribution.\footnote{We adopt $\Delta_{\rm m}=0.75$ throughout, as discussed in \S5.} $f_N$ is the 
normalization factor that matches the observed number density of the quasars in 
our sample ($\sim6.5 \times 10^{-6} \, h^3 {\rm Mpc}^{-3}$) with its expected 
value from the halo mass function and the assumed distribution of haloes. 
In the one-halo regime, the correlation function is 
measured using pairs of haloes that lie within the boundaries of individual haloes 
themselves. Therefore, in this regime, assumptions have to be made regarding the fraction of quasars that exist 
as a ``satellite'' of a central quasar ($f_{\rm sat}$), and regarding the fraction of quasars 
that reside at the center of dark matter haloes ($p$). 
The halo occupation model then has ``central'' and ``satellite'' components such 
that $\langle N_{\rm cen}(M)\rangle = (1-f_{\rm sat})\langle N(M)\rangle$ and $\langle 
N_{\rm sat}(M)\rangle = f_{\rm sat} \langle N(M)\rangle$ \citep{bw02,kra04, zh05}.
KO12 justifies ignoring the mass dependence of the satellite fraction 
(i.e., $f_{\rm sat}(M) \simeq f_{\rm sat}$) by choosing a relatively narrow ($\sim 2$ 
dex) range for the halo mass distribution around the characterized minimum mass, 
$M_m$. Similar to their approach, we assume a constant satellite fraction and 
investigate any difference by relaxing the width of the halo mass distribution, 
$\Delta_m$, as an additional fitting parameter in our HOD model. 

We further adopt an ``independent quasar activity'' assumption for the central 
and satellite quasars. This facilitates the use of Poisson statistics for the 
integrator in the description of the one-halo term of the power 
spectrum used in KO12 \citep[see also][]{sel00}:

 \begin{multline}
P_{1h}(k) = \frac{1}{n^{2}_{q}} \int \langle N(N-1)\rangle u(k,M)^p \frac{dn}{dM} dM  \\
\equiv 2 \langle N_{\rm cen} N_{\rm sat}\rangle u(k,M) + \langle N_{\rm sat}(N_{\rm sat}-1)\rangle |u(k,M)|^2 \\
= [2 f_{\rm sat}(1-f_{\rm sat}) u(k,M)+f_{\rm sat}^2 |u(k,M)|^2] \langle N(M)\rangle,
 \end{multline}

where $u(k,M)$ is the Fourier transform of the quasar number density profile in 
a dark matter halo of mass $M$. We adopt an NFW profile \citep{nfw97} with 
concentration parameter
\begin{equation}
c(M,z) = \frac{c_{0}}{1+z}\left[\frac{M}{M^{*}}\right]^{-0.13},
\end{equation}
 where $M^{*}$ is the nonlinear mass scale for collapse at $z=0$
 \citep[e.g.][]{cs02}. In this work, we initially chose $c_0=9$ and investigated 
the change in the best-fit parameters for a model with haloes with 10 times 
higher concentration (see the right panel of Fig. \ref{fig5:lumcomp}). A weak 
dependency of the HOD model on the concentration parameter is reported by a 
number of studies \citep{ric12,ric13,sh13} that used a different HOD formalism 
where the distribution of the central and satellite haloes are assumed to be a 
softened three-parameter step function and a two-parameter power-law 
respectively.
 We further discuss the differences between our chosen parameters and those 
previously used in the literature in $\S$\ref{res} where we summarize the 
differences between our assumptions and their potential effect on the best-fit 
parameters.    
\begin{table*}
\centering
\begin{tabular}{lccc}

\hline
\hline
Data &$\rm M_m (\times 10^{12}\, h^{-1} M_{\odot})$ &$f_{\rm sat}$& $\chi_{\rm red}^2$ (d.o.f)\\
\hline
 \vspace{.1in}
 4KDE+15eBOSS+8KO12& $1.45_{-0.24}^{+0.30}$& $0.024_{-0.007}^{+0.006}$& 1.46 (24)\\
  \vspace{.1in}
 10KO12+15eBOSS & $1.44_{-0.24}^{+0.30}$& $0.012_{-0.002}^{+0.003}$& 0.836 (22)\\
 \vspace{.1in}
 4KDE+15eBOSS & $2.31_{-0.38}^{+0.41}$& $0.071_{-0.009}^{+0.004}$& 0.919 (16)\\
 \vspace{.1in}
 4KDE+15eBOSS ($10\times \bar c$)& $3.42_{-0.48}^{+0.40}$& $0.065_{-0.007}^{+0.007}$& 1.019 (16)\\

\hline
\end{tabular}
\caption{ Summary of the fitting results using different data that participated 
in the fit. The second and third columns are the best-fit parameters and their 
1-$\sigma$ uncertainties. The last column lists the reduced $\chi^2$ (and the 
degrees of freedom).}\label{tab:res}
\end{table*}

\section{Results and Discussion}\label{res}

We fit our HOD model to three  
combinations of the measured $w_p$, as shown in 
Fig.\,\ref{fig3:obs} and summarized in Table\,\ref{tab:res}. 
On kpc-scales, we fit the modeled $w_p$ to combinations of the four bins of transformed $W_p$ to $w_p$ from the
``KDE-complete'' sample of \citet{ef17} and the eight bins from KO12 (these are denoted as ``KDE'' and KO12, respectively, in 
Fig.\,\ref{fig3:obs}). On Mpc-scales, we fit the modeled $w_p$
to the fifteen bins measured from eBOSS quasars with $g<20.85$. Measurements from KO12 provide
more coverage on kpc-scales than those of \citep{ef17} but with much 
higher uncertainty due to their smaller samples of close pairs (26 versus 44), particularly on
the smallest scales that might be expected to drive constraints on the satellite fraction. 

To constrain the shape and amplitude of the small-scale clustering of quasars with the full slate of
data, we also combine the KDE and KO12 measurements. 
The best-fit model to this ``full'' set of data is shown as the solid blue curve in Fig.\,\ref{fig4:bfit}. This
model has $\rm M_{m} = 1.45_{-0.24}^{+0.30} \times 10^{12}\, h^{-1} M_{\odot}$, and $f_{\rm 
sat} = 0.024_{-0.007}^{+0.006}$ with $\chi_{\rm red}^{2} = 1.46$. 
In assessing whether this ``full" model fit is truly meaningful, it is worth
considering the differences between our quasar samples and that of KO12. The quasar sample used in 
the clustering measurement and HOD analysis of KO12 matches the characteristics 
of our quasar sample reasonably well in number density and redshift (the sample of KO12
has $n \sim1.4 \times 10^{-6} \, h^3 {\rm Mpc}^{-3}$ and $\bar z\sim1.4$). 
Assuming that quasars do not evolve rapidly over the timescales that correspond 
to $\Delta z \sim 0.1$, it is reasonable to combine a slightly lower redshift 
measurement with the KDE measurements at $z \sim 1.5$. 

To investigate how each of the individual kpc-scale data sets affect 
the fitted model, we performed the MCMC fit separately to each of the 
KO12-plus-eBOSS and KDE-plus-eBOSS data sets. The 
green and red solid lines in Fig.\,\ref{fig4:bfit} depict the best-fit model for 
those two sets of measurements. Table\,\ref{tab:res} shows that the sole use of 
KO12 for the kpc section of the observed $w_p$ results in the lowest $M_{\rm min}$ 
and $f_{\rm sat}$ (green curve). On the other hand, neglecting the KO12 measurements 
in favour of the more precise, but scale-limited, KDE measurements, results in the highest values for the best-fit $M_{\rm min}$ and 
$f_{\rm sat}$ (red curve). We note that we only fit two parameters to our data as
$M_{\rm min}$ and $f_{\rm sat}$ appear to be the main drivers of HOD differences on small scales. We verified 
that relaxing the width of the halo mass distribution, $\Delta_{\rm m}$, as a 
third parameter has no significant effect on the shape of the best-fit model. 
Our best-fit value for $\Delta_{\rm m}$ always remains consistent with the value of
0.75 that was reported by KO12. Therefore, we have adapted a value of $\Delta_{\rm m}=0.75$ for all fits presented in this work.

We also examined fitting a model with $10\times$ 
higher NFW concentration parameter ($10 \times c$ in Eqn. 8) to our KDE+eBOSS data set. The 
solid orange curve in Figure\,\ref{fig4:bfit} depicts such a model, which
clearly has a steeper slope at kpc scales and shows more agreement with the 
observed higher clustering signal of the KDE data and the corresponding sharp drop 
to the eBOSS data on Mpc-scales. As also noted in KO12, models with a higher
concentration parameter are better able to reproduce the generally strong clustering of quasars
that is observed at kpc-scales. However, the best fit parameters for such a model (see 
Table\,\ref{tab:res}) do not show a significant improvement in the quality of the 
fit. Most importantly, the best-fit satellite fraction for a model with $10\times$ higher concentration parameter
is roughly the same as a model with a more typical concentration parameter 
($f_{\rm sat} \sim 0.065$ compared to $f_{\rm sat}\sim0.071$).  Comparing the fitting results in 
the first and third rows of Table\,\ref{tab:res} implies that including measurements from 
KO12 in our constraints results in a consistent set of 
best-fit parameters with that reported in KO12. Unsurprisingly, excluding 
data from K012 only affects the one-halo-term and results in a much larger 
satellite fraction in our work ($\sim 3 \times f_{\rm sat}$). This larger
satellite fraction is $\sim 1.5 \times f_{\rm sat}$ compared to what KO12 reports 
when using a similar NFW profile ($f_{\rm sat}\sim 0.071$
compared to $f_{\rm sat}\sim 0.048$).  

\begin{figure*}
    \centering
    \begin{subfigure}{
     \centering
       \includegraphics[angle=0,scale=0.5]{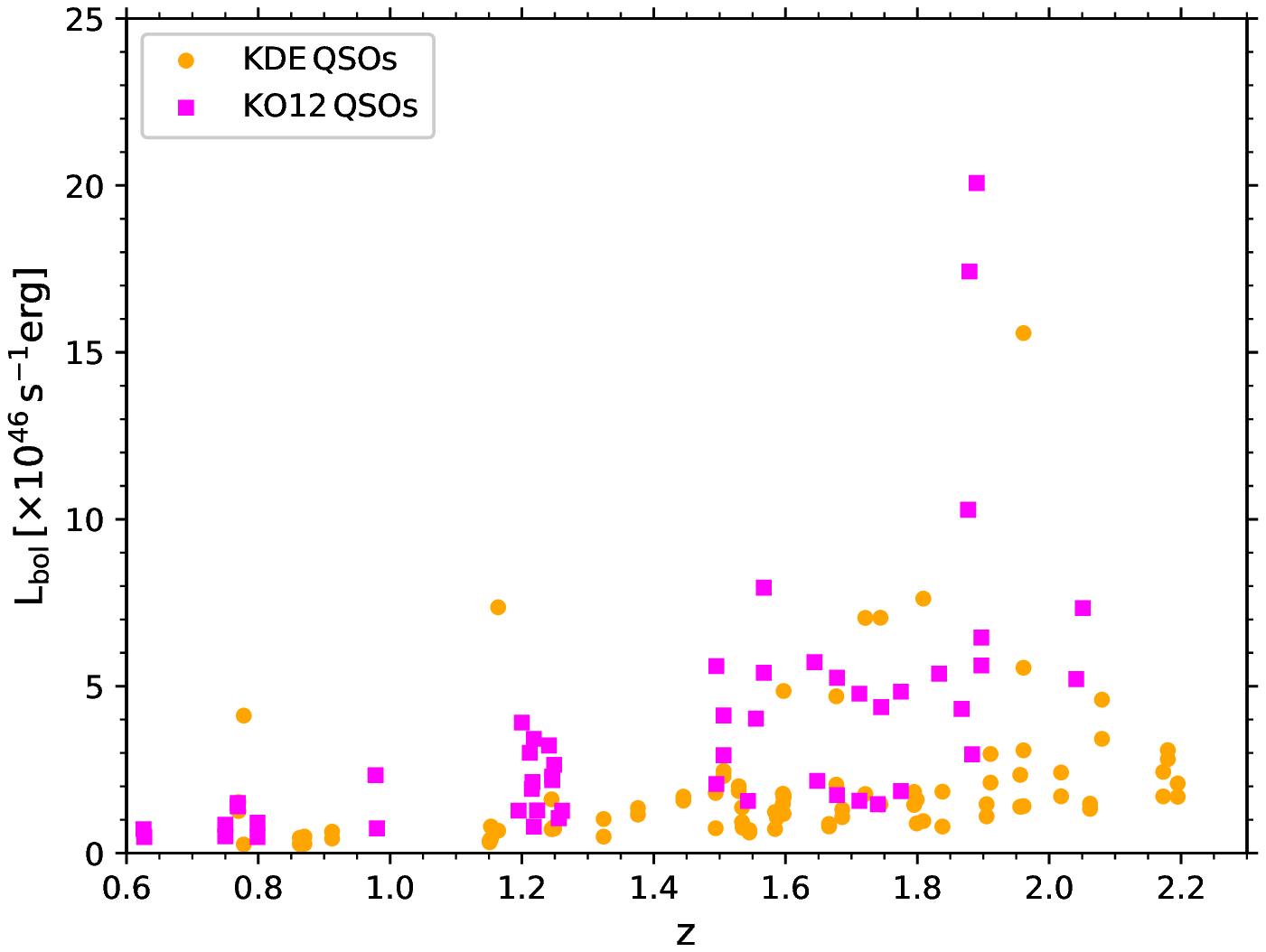}
       \includegraphics[angle=0,scale=0.5]{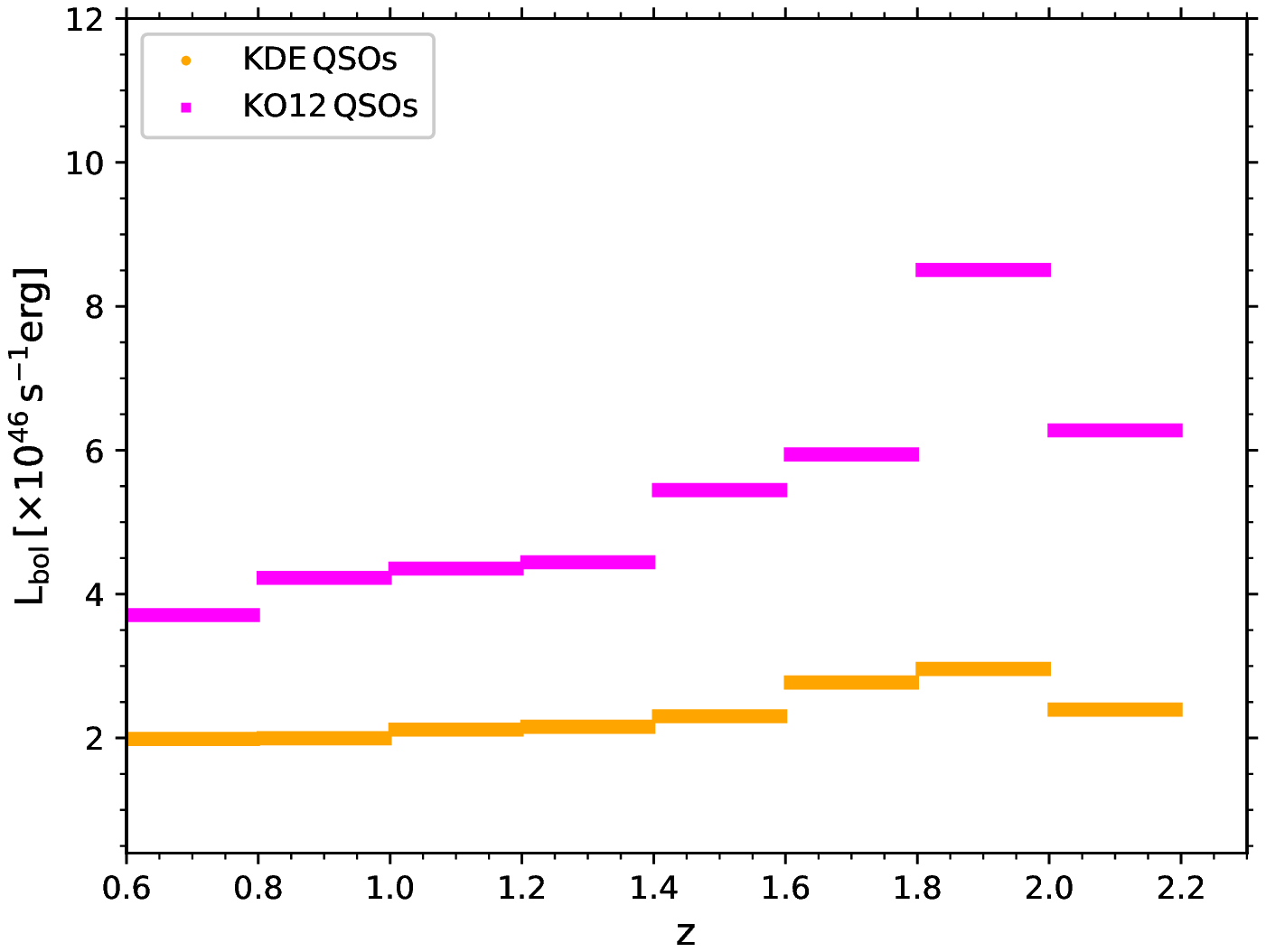}}
    \caption{Left: Bolometric luminosity of quasars in the KDE and KO12 samples 
for each member of a quasar pair (i.e.\ 88 luminosity values for the 44 pairs in the KDE 
sample and 52 luminosity values for the 26 pairs in the KO12 sample). Right: The 
average luminosity of the two quasar samples in bins of redshift.}\label{fig5:lumcomp}
    \end{subfigure}
\end{figure*}

One possible explanation for the significantly different satellite fractions we derive for the KDE and KO12
samples is that quasars of different luminosity populate individual dark
matter haloes in a different fashion, which could produce a pronounced luminosity-dependence
to quasar clustering on small scales. We can determine the typical luminosities of quasars in
the KDE and KO12 samples using the $i$-band magnitudes and redshifts of the
pair members reported in Table 5 of \cite{ef17} and Table 1 of KO12. 
We convert from apparent magnitude to bolometric luminosity using
Eqn.\,1 of \citet{she09a} and Eqn.\,1 of \citet{ric06}. 
Figure \ref{fig5:lumcomp} shows the difference in bolometric luminosity of the 
quasars in the KDE and KO12 samples. Evidently, KO12 quasars are 
somewhat more luminous than KDE quasars across the full redshift range of the 
two samples (averaging around $\sim 2 \times $ brighter in total).

Recent Mpc-scale clustering measurements for quasars at redshifts $0.9<z<3.6$ did not find compelling evidence for a strong 
luminosity dependence to quasar clustering \citep[see, e.g., ][for BOSS and 
eBOSS quasar samples]{ef15,la17}. But, recent HOD analyses suggest 
that local ($z \sim 0.01$--0.1), low-luminosity AGNs 
reside in similar dark matter haloes to that of galaxies with similar stellar 
mass \citep[see, e.g.,][]{pow18} and redder, brighter galaxies are known
to cluster more strongly \citep[e.g.][]{zeh05}. Further, the conditional luminosity 
function of redder, brighter galaxies implies that the clustering of central galaxies 
exceeds what would be extrapolated from the analytical form that fits the satellite population \citep[see, again,][]{zeh05}. 
Given that local, low-luminosity AGNs track galaxy clustering, and the statistics of central and
satellite galaxies depend on color and luminosity, it seems plausible
that there exists a luminosity threshold at which the luminosity-dependence of the
statistics of central and satellite quasars decouple. 
Ideally, hydrodynamic simulations would be used to study the luminosity functions
of central and satellite quasars. Modern simulations such as Illustris \citep{vog14} can constrain the evolution 
of satellite galaxies and low-luminosity AGN by studying merger trees \citep{rod15} 
for sub-haloes in cluster-like host dark matter haloes \citep[see, e.g.,][]{nie18}. 
However, it is possible that high-luminosity quasars trace the small galaxy group environments 
in which major mergers are most efficient \citep[e.g.][]{hop08}, which would make quasar
statistics harder to resolve in simulations than for AGN and galaxy statistics at the cluster-scale. Nevertheless,
the strong luminosity dependence for the satellite fraction of quasars implied by our study should ultimately
be studied in higher-resolution, physically motivated simulations.

\section{Summary and Conclusions}\label{con}

In this paper, we have studied constraints from quasar clustering measurements over 4 orders of magnitude in comoving scale. On small scales ($43.3 \lesssim r_p \lesssim 92.3\,h^{-1}\,{\rm kpc}$) our sample consists of the ``KDE-complete`` sample of 44 $g<20.85$ quasar pairs presented in \citet{ef17}. On moderate scales, we use
the 26 $i \lesssim 19.1$ quasar pairs presented by \citet[][denoted as KO12 throughout the paper]{ko12}. 
On large scales ($1 \lesssim r_p \lesssim 100\,h^{-1}\,{\rm Mpc}$ we use 40{,}821 $g<20.85$ quasars drawn from the extended Baryon Oscillation Spectroscopic Survey \citep[eBOSS;][]{daw16}. Our ``KDE-complete`` and eBOSS samples are well-matched in number density, redshift 
range and magnitude. 

We derived the two-point correlation function ($w_p$) for our samples and fit $w_p$ in the context of the Halo Occupation Distribution (HOD) in 
order to estimate the satellite fraction and average minimum halo mass of the dark matter haloes that host quasars. 
To construct our HOD model, we used a ``Gaussian-like'' functional form for the mean number of quasars that inhabit each halo ($\langle 
N(M)\rangle$), consistent with the idea that quasars with a wide range of luminosity are hosted by dark matter haloes with a narrow range of mass
\citep[e.g., ][]{Lid06}. Our model also assumes that central and satellite quasars are independent and that the number density of quasars
in a dark matter halo can be modeled by an NFW profile.

We found that the satellite fraction produced by our HOD model changed significantly depending on whether we use our ``KDE-complete``
data to constrain small-scale quasar clustering, or the data from KO12 (see Fig. \ref{fig4:bfit} and Table \ref{tab:res}). This difference in
satellite fraction could also be explained by a different concentration parameter for the NFW profile driving the one-halo term, such that
the ``KDE-complete`` sample inhabits much more concentrated haloes. Assuming that there is no significant evolution 
between the average redshift of the KO12 and KDE-complete samples ($z\sim1.4$ and $z\sim1.5$, or a difference of $\sim$260\,Myrs), we attribute
differences in the clustering to differences in luminosity, as the KO12 sample is several times more luminous than the KDE-complete sample.

We find that the one-halo term used in HOD models is particularly flexible, and can be sensitive to satellite fraction, density profile, and the functional form
of $\langle N(M)\rangle$. Therefore, the most physically motivated way to test our proposed luminosity dependence to quasar clustering on small
scales may be via physically motivated hydrodynamic simulations \citep[e.g.,][]{ar18,bh18c}. The coming era of large quasar samples and high-resolution simulations
should ultimately be able to test for any luminosity dependence to quasar clustering on small scales. In particular, the host halo mass dependence 
of the satellite fraction, and the conditional luminosity function of satellite quasars, remain open avenues for investigation.

\section*{Acknowledgments}
SE and ADM acknowledge support by the National Science Foundation through  
grant number 1616168. SE was also supported by the U.S. Department of Energy, Office of Science, through award number DE-SC0018044
and ADM was also supported by the U.S. Department of Energy, Office of Science, Office of High Energy Physics, under
Award Number DE-SC0019022. SE thanks Rita Tojeiro and Ashley Ross for useful discussions.

Funding for the Sloan Digital Sky Survey IV has been provided by the Alfred P. 
Sloan Foundation, the U.S. Department of Energy Office of Science, and the 
Participating Institutions. SDSS acknowledges support and resources from the 
Center for High-Performance Computing at the University of Utah. The SDSS web 
site is \url{http://www.sdss.org}.

SDSS IV is managed by the Astrophysical Research Consortium for the Participating 
Institutions of the SDSS Collaboration including the Brazilian Participation 
Group, the Carnegie Institution for Science, Carnegie Mellon University, the 
Chilean Participation Group, the French Participation Group, Harvard-Smithsonian 
Center for Astrophysics, Instituto de Astrof\'{i}sica de Canarias, The Johns 
Hopkins University, Kavli Institute for the Physics and Mathematics of the 
Universe (IPMU) / University of Tokyo, Lawrence Berkeley National Laboratory, 
Leibniz Institut f\"{u}r Astrophysik Potsdam (AIP), Max-Planck-Institut f\"{u}r 
Astronomie (MPIA Heidelberg), Max-Planck-Institut f\"{u}r Astrophysik (MPA 
Garching), Max-Planck-Institut f\"{u}r Extraterrestrische Physik (MPE), National 
Astronomical Observatories of China, New Mexico State University, New York 
University, University of Notre Dame, Observat\'{o}rio Nacional / MCTI, The Ohio 
State University, Pennsylvania State University, Shanghai Astronomical 
Observatory, United Kingdom Participation Group, Universidad Nacional 
Aut\'{o}noma de M\'{e}xico, University of Arizona, University of Colorado 
Boulder, University of Oxford, University of Portsmouth, University of Utah, 
University of Virginia, University of Washington, University of Wisconsin, 
Vanderbilt University, and Yale University.
\newpage
\bibliographystyle{mn2e.bst}
\bibliography{hodrefs}
\end{document}